# Trans-Neptunian objects: a laboratory for testing the existence of non-Newtonian component of the gravitational force

Dragan Slavkov Hajdukovic[1]
PH Division CERN
CH-1211 Geneva 23
dragan.hajdukovic@cern.ch
[1]On leave from Cetinje, Montenegro

Abstract

It is not known if, in addition to the Newton's inverse square law component, the gravitational force has some non-Newtonian, long-range components that have escaped detection until now. For example, the non-Newtonian component of the gravitational force naturally arises if gravity is interpreted as an entropic force, or under far reaching hypothesis that quantum vacuum contains virtual gravitational dipoles. We point out that some trans-Neptunian objects (for instance a binary system or a dwarf planet with its satellite) might be a good laboratory to establish the eventual existence of non-Newtonian components of gravity. The key points are that, in the case of an ideal two-body system, the perihelion precession can be caused *only* by a gravitational force that deviates from the inverse square law and that the perihelion precession rate is larger in systems with smaller mass. It is shown, that in some trans-Neptunian (two-body) systems, the perihelion precession rate caused by internal interactions might be larger than the (inevitable) precession induced by external gravitational field.

1. Introduction

According to the classical (Newtonian) celestial mechanics, the orbit of a planet can be an ellipse fixed with respect to the Sun, only if: (a) the gravitational field of the Sun has a perfect spherical symmetry, (b) there are only two bodies (Sun and the planet), i.e. there are no additional bodies that perturb the spherical symmetry, (c) the gravitational force strictly follows the Newton inverse square law. Any departure from the spherical symmetry and/or the inverse square law of gravity, leads to the precession of the perihelion. Of course, the Solar System is a many-body system and the Sun is not perfectly spherical; consequently the perihelion precession is a common propriety of all elliptical orbits (orbits of planets around the Sun, orbits of satellites around the planets and the mutual orbits in binary systems).

The perihelion precession predicted by classical theory (Newtonian mechanics together with the inverse square law for gravity) is close to the observed values; the largest discrepancy occurs for the Mercury (Mercury's orbit process at a rate that is about $8\%$ greater than the predicted one). The discrepancies have been explained by the General Relativity (see for instance Fitzpatrick 2012) which adds to the Newtonian result the following term:

$$\Delta\omega_{gr} = \frac{6\pi G}{c^2(1-e^2)}\frac{M_S}{a} \qquad (1)$$

where $\Delta\omega_{gr}$ is the extra rotation per orbit in radians, $M_S$ the mass of the Sun, $a$ the semi-major axis of the orbit and $e$ the eccentricity of the ellipse. The general relativistic correction (1) decreases when the semi-major axis increases, what is easy to understand because the relativistic effects are larger in a stronger gravitational field. For the Mercury, the relation (1) predicts the additional *43 arc seconds per century*, and historically it was the first success of General Relativity. For the Neptune the correction (1) is only *0.0008 arc seconds per century*, what is five orders of magnitude smaller than the classical contribution; hence for the trans-Neptunian objects, the general relativistic correction (1) can be completely neglected.

The starting point of our paper is that the precession of the perihelion in the case of an ideal two-body system can exist *only if* the gravitational force has a non-Newtonian component (i.e. a component that deviates from the inverse square law). Hence, the precession in an ideal two-body system is the exclusive signature of non-Newtonian component of the gravitational force.

Of course the ideal (i.e. completely isolated) two-body system does not exist; the other bodies (i.e. the external gravitational field) are the inevitable and well established source of the perihelion precession. Consequently, the hypothetical precession, caused by the non-Newtonian interaction within the two-body system, would be always mixed with the precession induced by external objects. The best "laboratory" would be a system in which the internal source of precession dominates the external one.

Let us imagine that the system Earth-Moon can be transported far from the Sun; for instance, at a distance of 100AU, the external gravitational field would be about $10^4$ weaker, while the internal interactions would not be subject of any dramatic change. In principle, for every system, there is a critical distance from the Sun, so that for the greater distances, the internal properties of the system are dominant. The obvious idea is that it can be the case for some trans-Neptunian systems.

As example we will consider the following trans-Neptunian objects:
(a) Three dwarf-planets; Pluto with its satellite Sharon, Haumea with its satellites Hi'iaka and Namaka (Ragozzine and Brown, 2009) and Eris with its satellite Dysnomia (Brown, 2005, 2006, 2007)
(b) Two minor planets Quaoar with the satellite Weywot (Fraser and Brown 2010, Fraser et al. 2012) and Logos with the satellite Zoe (Grundy et al. 2011)
(c) The binary (66652) 1999 RZ$_{253}$ (Noll et al 2004)

Table 1 contains the basic information for all these systems; for comparison we have also included the system Earth-Moon.

Table 1: The main characteristics of the considered systems

|  | $a_{sun}$ [AU] | $T_{sun}$ [years] | Mass [kg] | Diameter [km] | a [km] | Eccentricity | Period [days] |
|---|---|---|---|---|---|---|---|
| Pluto | 39.3 | 246 | $1.3 \times 10^{22}$ | 2,306 |  |  |  |
| Sharon |  |  | $1.5 \times 10^{21}$ | 1,206 | $1.75 \times 10^4$ | 0.0022 | 6.39 |
| Haumea | 43.1 | 283.3 | $4 \times 10^{21}$ | 1400 |  |  |  |
| Hi'iaka |  |  | $1.8 \times 10^{19}$ | 340 | $5 \times 10^4$ | 0.051 | 49.5 |
| Haumea | 43.1 | 283.3 | $4 \times 10^{21}$ | 1400 |  |  |  |
| Namaka |  |  | $1.8 \times 10^{18}$ | 170 | $2.5 \times 10^4$ | 0.249 | 18.3 |
| Eris | 68 | 560.9 | $1.67 \times 10^{22}$ | 2300 |  |  |  |
| Dysnomia |  |  | $\sim 10^{19}$ | 340 | $3.74 \times 10^4$ | e<0.013 | 15.77 |
| Quaoar | 43.4 | 286 | $1.6 \times 10^{21}$ | 1170 |  |  |  |
| Weywot |  |  | $\sim 10^{17}$ | 74 | $1.45 \times 10^4$ | 0.14 | 12.44 |
| Logos | 45.1 | 302.8 | $2.7 \times 10^{17}$ | 80 |  |  |  |
| Zoe |  |  | $1.5 \times 10^{17}$ | 66 | 8,200 | 0.546 | 309.9 |
| (66652) | 43.8 | 290 | $2.4 \times 10^{18}$ | 166 |  |  |  |
| Borasisi |  |  | $1.34 \times 10^{18}$ | 137 | 4,660 | 0.47 | 46.3 |
| Earth | 1 | 1 | $6 \times 10^{24}$ | 6,371 |  |  |  |
| Moon |  |  | $7.3 \times 10^{22}$ | 3,470 | $3.84 \times 10^5$ | 0.055 | 27.3 |

Note: $a_{sun}$ and $T_{sun}$ denote the semi-major axis and the period of the orbit of the system around the Sun; other quantities are internal characteristics of the system

In Section 2 we consider ideal two-body systems. In Section 3 we study real systems presented in Table 1, combining the results of Section 2, with the effects of the external gravitational field. Section 4 is devoted to comments.

2. The perihelion precession in an ideal two-body system

Let us underline that the title of this section has no sense without assumption of a non-Newtonian component of gravity. Obviously, if such a component exists, it must be much smaller than the Newtonian component and hence can be treated as a small perturbation.

Let us consider an ideal two-body system with masses $M$ and $m$ and let us denote by $\vec{g}_a(r,\theta,z)$ the anomalous (non-Newtonian) gravitational acceleration; here $(r,\theta,z)$ are the cylindrical coordinates and $(r,\theta)$ the polar coordinates in the orbital plane. If $\vec{r}_0$, $\vec{\theta}_0$ and $\vec{z}_0$ denote the unit vectors, the anomalous acceleration may be written as

$$\vec{g}_a(r,\theta,z) = \overline{R}\vec{r}_0 + \overline{T}\vec{\theta}_0 + \overline{N}\vec{z}_0 \tag{2}$$

where $\overline{R}$, $\overline{T}$ and $\overline{N}$ are the magnitudes of the radial, tangential and normal component of the acceleration respectively.

As known from the classical celestial mechanics (see for instance the book of Murray and Dermott, 1999)

$$\frac{d\omega}{dt} = \frac{\sqrt{1-e^2}}{e}\sqrt{\frac{a}{G(M+m)}}\left[-\overline{R}\cos f + \overline{T}\sin f \frac{2+e\cos f}{1+e\cos f}\right] - \dot{\Omega}\cos I \tag{3}$$

In equation (3), $f$ denotes the true anomaly, $I$ is inclination of the orbit, $\Omega$ the longitude of ascending node and $\dot{\Omega} \equiv d\Omega/dt$ (it is important to know that $\dot{\Omega}$ is proportional to $\overline{N}$). The equation (3) is valid for *all perturbations (2) independently of their nature* and, for a given $\vec{g}_a(r,\theta,z)$, it allows to calculate the precession rate per a certain period of time.

In the simplest case of the spherical symmetry $\overline{T} = 0$, $\overline{N} = 0$ and $\dot{\Omega} = 0$. Consequently the equation (3) reduces to

$$\frac{d\omega}{dt} = -\frac{\sqrt{1-e^2}}{e}\sqrt{\frac{a}{G(M+m)}}\overline{R}(r)\cos f \tag{4}$$

In order to integrate the equation (4) it is necessary to know the function $\overline{R}(r)$ and to express $r$, $\cos f$ and $dt$ as functions of the eccentric anomaly $E$:

$$r = a(1-e\cos E) \tag{5}$$

$$\cos f = \frac{\cos E - e}{1 - e\cos E} \tag{6}$$

$$dt = \frac{1 - e\cos E}{n}dE \tag{7}$$

where $n \equiv 2\pi/T$ denotes "average" angular velocity (or the mean motion) and $T$ is the orbital period.

## 2.1 The case of a constant radial acceleration

Mathematically, the simplest possibility is the case of a constant radial perturbation $\overline{R}(r)$ in the equation (4). This mathematical simplification might correspond to a real physical phenomenon, as argued in the emerging theory that considers dark matter and dark energy as a consequence of the quantum vacuum containing the hypothetical virtual gravitational dipoles (Hajdukovic, 2010, 2011a, 2011b, 2012a, 2012b, 2012c). However, independently of any specific correction to the Newton law (for instance power-law entropic correction and Yukawa type corrections), the constant radial acceleration $\overline{R}(r)$ is a universal upper bound for all of them.

Using relations (6) and (7) a simple integration of the equation (4) gives the perihelion shift over the time interval $\Delta t$

$$\Delta\omega_{nN} = A_r\sqrt{1-e^2}\sqrt{\frac{a}{G(M+m)}}\Delta t \tag{8}$$

where $A_r$ denotes the constant value of $\overline{R}(r)$. In $\Delta\omega_{nN}$ we have used the subscript $nN$ to underline that that the source of precession is non-Newtonian component of the gravitational force.

It is evident from the result (8) that the perihelion precession rate is larger for systems with small mass $M+m$. Because of the large Solar mass it would be a tiny effect for orbits of planets around the Sun, but might be a few orders of magnitude larger for orbits of satellites around dwarf-planets or in small binary systems that are frequent in the trans-Neptunian "family".

What would be measured in real astronomical observations is the perihelion shift over a large time interval $\Delta t$; once this quantity is measured, the equation (8) allows calculating $A_r$. However, in order to see what the expected perihelion shifts are, we have calculated perihelion shifts for a tiny radial perturbation $A_r = 5\times 10^{-12}\, m/s^2$. The results are presented in the second column of the table 2.

Table 2: Comparison of the perihelion precession rate caused by non-Newtonian and Newtonian forces for $A_r = 5\times 10^{-12}\, m/s^2$

|  | $\Delta\omega_{nN}$ ["/century] | $\Delta\omega_N$ ["/century] | Comments |
|---|---|---|---|
| Pluto Sharon | 14 | 28 | Too small eccentricity |
| Haumea Hi'iaka | 44 | 163 | There is additional Newtonian precession because of ellipsoidal shape |
| Haumea Namaka | 31 | 60 |  |
| Eris Dysnomia | 19 | 13 | Eccentricity still not known |
| Quaoar Weywot | 38 | 40 | Weywot is too small for successful measurement |
| Logos Zoe | 1600 | 900 | Dominated by external field |
| (66652) Borasisi | 400 | 145 |  |
| Earth Moon | 3.2 | 7×10$^6$ | Dominated by external field |

We have used a very small perturbation $A_r$ (24 times smaller than the fundamental acceleration postulated in MOND and 175 times smaller than the acceleration involved in the Pioneer anomaly). In spite of such a miniscule value, the numbers in the second column of the Table 2, suggest that, with the fast improvement of technology, the measurements might be possible in the near future.

The important conclusion from the Table 2 is that (if considered as an ideal two body system) the system Earth-Moon has a small perihelion precession rate of about 3 arc seconds per century. Hence we have to look for systems with a mass much smaller than the total mass of the Earth and Moon; good "laboratories" should not have a mass larger than $\approx 10^{22}\, kg$.

In the real system Earth-Moon, the perihelion precession rate is $\approx 7\times 10^6\, arc\sec/century$, 6 orders of magnitude larger than in an isolated system, what is a consequence of the strong influence of the Sun. It is easy to calculate that, at the position of the Moon, the Newtonian gravitational field of the Sun $g_S \approx 6\times 10^{-3}\, m/s^2$ is larger than the gravitational field of the Earth $g_E \approx 2.7\times 10^{-3}\, m/s^2$. Because of the possibility that Newtonian and non-Newtonian components of gravity are related, it might happen that the external non-Newtonian component dominates, making impossible the use of relation (8). Hence the system Earth-Moon leads to the additional conclusion that we have to look for trans-Neptunian objects in which the internal Newtonian field is stronger than the external field.

The fact that a system is trans-Neptunian is not the guarantee that internal Newton's gravitation is stronger than the external one. The example is the Logos-Zoe system.

Some apparently good laboratories (like Quaor-Weywot) suffer from the small size of the satellite; even "detecting" Weywot is quite hard with current technology. Contrary to Weywot (a miniscule satellite with significant eccentricity of the orbit), Sharon is a large satellite but with nearly zero eccentricity, making once again the measurement impossible with the available technology. While the eccentricity of the orbit of Dysnomia is not known, the system Eris-Dysnomia may suffer from the same problem as Pluto-Sharon.

The dwarf-planet Haumea, illustrates the other kind of problem; it is an ellipsoid that deviates a lot from the spherical shape what is an internal source of the Newtonian perihelion precession that is not included in the Table 2.

### 2.2 Some other non-Newtonian components of gravity

Of course, in the general case we must allow possibility that the non-Newtonian component is not a constant. Probably the most studied is the Yukawa-type component of gravity (Adelberger et al. 2003, Merkowitz, 2010), with the non-Newtonian gravitational potential

$$V_Y(r) = -\frac{GMm}{r} \alpha \exp\left(-\frac{r}{\lambda}\right) \tag{9}$$

where $\alpha$ is the dimensionless strength and $\lambda$ is the length scale. Of course the non-Newtonian perihelion precession is again described by equation (4), but, according to (9) the radial perturbation $\bar{R}(r)$ is given by

$$\bar{R}(r) \equiv \bar{R}_Y(r) = -\alpha\left(1 + \frac{r}{\lambda}\right)\exp\left(-\frac{r}{\lambda}\right)\frac{GM}{r^2} \tag{10}$$

The other proposal that has attracted a lot of attention is entropic corrections to Newton's law that naturally appear if gravity is interpreted as an entopic force (Verlinde, 2011). In the simplest case (Modesto and Randono, 2010) the radial perturbation can be reduced to

$$\bar{R}(r) = 12\sqrt{\pi}\,\frac{b}{L_P}\frac{GM}{r} \tag{11}$$

where $b$ is a dimensionless parameter and $L_P$ the Planck length. The equation (4) with the perturbation (10) relates the perihelion shift to Yukawa's parameters $\alpha$ and $\lambda$, while with the perturbation (11) the shift is related to entropic parameter $b$. The measurement of shifts limits the possible values of parameters

### 3. Two body system in an external gravitational field

As already pointed out, the *unknown* precession caused by internal non-Newtonian interaction is always mixed with the precession induced by external (Newtonian) gravitational field. Fortunately, for a system that orbits around the Sun with a period $T_{sun}$, the Newtonian perihelion shift $\Delta\omega_N$ can be well approximated with (Murray and Dermott 1999, Urbassek 2009)

$$\Delta\omega_N \approx \frac{3\pi}{2}\frac{T_s}{T_{sun}^2}\Delta t \tag{12}$$

where $T_S$ denotes the internal period of the satellite within the system.

The third column of the Table 2 gives the Newtonian perihelion shifts. The eventual discovery of a perihelion precession rate, sharply different from the calculated one in the framework of the Newtonian mechanics, would be a strong sign of non-Newtonian component of the gravitational force.

4. Comments

First, let us underline that the ultimate possibility, allowing much higher precision, would be to study the orbit of an artificial satellite of a minor planet; of course if the future scientific considerations justify such a complicated and expensive experiment. In fact, humanity has already entered such kind of missions; the well-known example is New Horizons, a NASA robotic spacecraft mission currently en route to the dwarf planet Pluto.

Second, as suggested in the recent publications, the constant non-Newtonian radial acceleration might be a real physical phenomenon assuming that the quantum vacuum contains virtual gravitational dipoles. The hypothesis of the virtual gravitational dipoles might be the bases to understand the nature of dark matter and dark energy (Hajdukovic, 2010, 2011a, 2011b, 2012a, 2012b, 2012c).

In fact, if the quantum vacuum contains the virtual gravitational dipoles, the gravitational field of a body (Sun, Earth…) immersed in the quantum vacuum, should produce vacuum polarization characterized with a gravitational polarization density $\vec{P}_g(r)$ (i.e. the gravitational dipole moment per unit volume). In principle the space around a spherical body can be divided in two regions. In the inner region, up to a critical distance, the gravitational field is sufficiently strong to align all dipoles along the field; consequently $P_g(r)$ has a constant value (in fact a maximum value) that may be denoted $P_{g\,max}$. If $P_{g\,max}$ exists it should be considered as a fundamental constant of the quantum vacuum.

The key point is that the constant perturbation $A_r$ in the equation (8) is proportional to $P_{g\,max}$, i.e.

$$A_r \equiv -4\pi G P_{g\,max} \tag{13}$$

If the hypothesis of the gravitational dipoles is correct, the trans-Neptunian objects have potential to determine the fundamental constant $P_{g\,max}$ related to the gravitational properties of the quantum vacuum and to be a necessary complement to the forthcoming experiments at CERN (Kellerbauer et al. 2012, Perez and Sacquin 2012).